# In-situ Piezoresponse Force Microscopy Cantilever Mode Shape Profiling

R. Proksch[1]

*Asylum Research, an Oxford Instruments Company, Santa Barbara, CA USA*

The frequency-dependent amplitude and phase in piezoresponse force microscopy (PFM) measurements are shown to be a consequence of the Euler-Bernoulli (EB) dynamics of atomic force microscope (AFM) cantilever beams used to make the measurements. Changes in the cantilever mode shape as a function of changes in the boundary conditions determine the sensitivity of cantilevers to forces between the tip and the sample. Conventional PFM and AFM measurements are made with the motion of the cantilever measured at one optical beam detector (OBD) spot location. A single OBD spot location provides a limited picture of the total cantilever motion and in fact, experimentally observed cantilever amplitude and phase are shown to be strongly dependent on the OBD spot position for many measurements. In this work, the commonly observed frequency dependence of PFM response is explained through experimental measurements and analytic theoretical EB modeling of the PFM response as a function of both frequency and OBD spot location on a periodically poled lithium niobate (PPLN) sample. One notable conclusion is that a common choice of OBD spot location – at or near the tip of the cantilever – is particularly vulnerable to frequency dependent amplitude and phase variations stemming from dynamics of the cantilever sensor rather than from the piezoresponse of the sample.

---

[1] roger.proksch@oxinst.com

**I. INTRODUCTION: Electromechanics**

Coupling between electrical and mechanical phenomena is a common characteristic of many natural systems, with examples ranging from piezoelectric and ferroelectric[1, 2, 3, 4, 5] and energy storage materials[6, 7, 8] to biological samples[9, 10] including molecules and complex tissues.[11] Based on the atomic force microscope (AFM),[12] piezoresponse force microscopy (PFM)[13, 14] has become an established, powerful tool for functional nanoscale imaging, spectroscopy, and manipulation of electromechanical materials. Another technique that relies on an oscillating tip-sample bias is the recently developed electrochemical strain microscopy (ESM).[15, 16] In ESM, the oscillating tip-sample bias induces localized ionic motion that in turn causes a strain that is coupled to the cantilever through the tip.

In the following, we will first discuss conventional PFM, noting the challenges there have been in obtaining repeatable, quantitative measurements of the inverse piezo sensitivity. These challenges also exist for more generalized to electromechanical measurements where there are other coupling mechanisms between a bias applied to a tip and a resulting strain in the material. We then make detailed specific experimental observations on a common sample – lithium niobate, noting the strong frequency dependence and deviation from ideal behavior. This leads to an investigation of the effect of the optical beam detector (OBD) spot location on the measured observables. The resulting spectrograms – measurements of the frequency-dependent amplitude and phase response of the cantilever versus OBD spot position – reveal a number of interesting and surprising results. The experimental spectrograms are shown to explain the DC voltage-dependent response. They also in turn suggest limits for quantitative operation of PFM and ESM, at least in the context of OBD based measurements.

**A. Piezoresponse Force Microscopy**

PFM is based on the converse piezoelectric effect, with the cantilever in contact mode while the tip-sample voltage is modulated with a periodic tip bias $V_{tip} = V_{DC} + V_{AC} \cos(\omega t)$, where the drive frequency $\omega$ is chosen to be well above the feedback bandwidth, roughly 5-7 kHz in this case. This drive generates an oscillating electric field below the tip that causes localized deformations in the sample surface, which in turn acts as a mechanical drive for the cantilever. The piezoelectric response of the surface is detected as the first harmonic component $A_{1\omega}$ of the tip deflection amplitude $A = A_0 + A_{1\omega}\cos(\omega t + \phi)$. When the response is dominated by the piezoelectric contribution, the phase $\phi$ of the electromechanical response of the surface is dependent on the polarization direction. As mentioned above, there are other forces present that also respond at this first harmonic, including localized electrostatic forces between the *tip* and sample surface charge and delocalized (long-range) electrostatic forces between the *body* of the cantilever and the sample surface charge. In general, the larger the piezoresponse amplitude relative to other contributions, the better the PFM contrast and quantitative interpretation of the data.[17, 18] Over the past years, a number of approaches for maximizing the PFM response and minimizing or eliminating the electrostatic components have been developed;[19, 20] however, it still remains a significant challenge.

Typically by using a lock-in amplifier, the amplitude and phase of these electromechanical vibrations at the first drive harmonic are extracted and provide information on the piezoelectric orientation and mechanical properties (stiffness and dissipation) of the material in the vicinity of the tip. Explicitly, the amplitude is related to the functional sensitivity of the material, and the phase is related to the domain orientation. The cantilever resonance can be used to improve the signal to noise of these measurements with resonance-tracking techniques such as the lock-in-based Dual AC Resonance Tracking (DART)[21] method or the adaptive digital-synthesizing Band Excitation (BE) method.[22]

**II. Quantifying AFM Electromechanical Response**

One of the ongoing challenges of PFM, ESM, and related techniques is the accurate characterization of functional parameters. In the case of PFM, the most common functional parameter is the inverse piezo sensitivity, a measure of the strain response of a piezoelectric material to an applied voltage, typically quoted in units of nm/V. Issues with accurate measurement of this parameter include

(i) the tensorial nature of electromechanical coupling, (ii) uncertainties in the tip-sample mechanical interface, (iii) uncertainties in the calibration of the mechanical and OBD sensitivity (defined below) of the cantilever and (iv) forces between the body of the cantilever and tip of the cantilever competing with the piezoelectric actuation.[23]

Following Jesse et al.,[24] it is common to define an "effective" inverse piezo sensitivity $d_{\text{eff}}$ by $d_{\text{eff}} = A_{em}/V_{tip}$, where $A_{em}$ is the amplitude of the electromechanical strain driving the cantilever tip and $V_{tip}$ is the applied voltage. This sensitivity combines the components of the piezoelectric tensor to describe the resulting response of the PFM cantilever to the applied voltage along the z-axis.[25, 26, 27] The response $A_{1\omega}$ of the cantilever at the first harmonic of the AC drive voltage is given by a combination of localized piezoresponse, localized electrostatic interactions between the *tip* and the sample and long-range electrostatics interactions between the sample and the cantilever *body*. There are similar relationships for lateral components, however here we will only consider the vertical PFM response. The electrostatic coupling is generally responsible for a background signal at the drive frequency that complicates interpretation of the PFM response.[28, 29] "Ideal" PFM measurements of ferroelectric materials should follow some simple trends:

1. Measurements are frequency-independent, at least below the first contact resonance frequency.
2. The amplitude should be independent of the ferroelectric polarization direction.[30]
3. The phase shift across oppositely polarized domains should be 180°.

To elaborate on point #1 above, the mechanical resonances of bulk ferroelectric crystals are well above the 1 kHz – 1 MHz PFM range of operation.[31] However, it is well known that the drive frequency of the electrical excitation can have a profound effect on the measured PFM signal,[32, 33] with at least part of the frequency dependence due to the contact resonance of the cantilever.[34] Since the contact resonance frequency depends on the contact stiffness of the tip-sample interaction, as the contact area changes while the cantilever scans over the surface, operating too close to the cantilever resonance can lead to artifacts in the response. Harnagea et al.[35] suggested operating the contact cantilever near, but not exactly on a resonant frequency to enhance the response and at the same time avoid artifacts associated with changes in the contact resonance due to adhesion effects.

In addition to electrostatic coupling, other sources of background can cause crosstalk in the PFM response, including instrumental resonances and/or electrical background signals. The dangers of exciting instrumental electrical or mechanical resonances in the AFM while making PFM measurements have been elaborated already.[36, 37] Briefly, in a poorly designed excitation system, unwanted electrical couplings in the conductive path between the electrical oscillator and the cantilever tip can, for example also drive the "shake piezo", the device commonly used for mechanical actuation of the cantilever in tapping mode or other measurements. If this occurs, excitation of the shake piezo appears as cantilever motion, indistinguishable from motion originating from the sample electromechanical strain. Similarly, these same drive voltages can couple into a poorly isolated photodetector circuit, leading to a false background at the first harmonic, affecting both the amplitude and phase of the measured signal. This coupling often has an ill-defined frequency dependence leading to further difficulties in interpreting or removing the background. In the Cypher AFM used here, these effects have been effectively eliminated through careful design of the electrical signal routing and shielding.

In most cases, single-frequency PFM has been limited to a few hundred kHz or lower[38] with some exceptions.[39, 40] There are potential advantages to operation at higher frequencies including improved signal to noise. However, as mentioned above there are also complications associated with the cantilever contact resonance, because local elastic and contact area variations due to topography can lead to crosstalk.[24] Crosstalk issues with on-resonance operation have been improved through resonance-tracking techniques such as the DART and BE approaches mentioned above. Both of these techniques allow the determination of the driving force or strain in PFM by properly accounting for shifting resonance and quality factor as the cantilever scans over the surface. These measurements are analyzed in terms of the simple harmonic oscillator (SHO) model, where the amplitude is described by the well-known expression

$$A = \frac{k_{cant} A_{em} \omega_0^2}{\sqrt{(\omega_0^2 - \omega^2)^2 + (Q/\omega\omega_0)^2}} = \frac{d_{eff} V_{tip} k_{cant} \omega_0^2}{\sqrt{(\omega_0^2 - \omega^2)^2 + (Q/\omega\omega_0)^2}}. \tag{1}$$

A related expression exists for the SHO phase but is not addressed here. In Eq. (1), the oscillator resonance and drive frequencies are $\omega_0 = 2\pi f_0$ and $\omega = 2\pi f$ respectively. The dimensionless quality factor $Q$ of the SHO ranges from approximately 30 to 200 for a wide range of silicon cantilevers used in electromechanical measurements made in ambient conditions. $k_{cant}$ is the cantilever spring constant, typically taken as the end-loaded stiffness, which ranges from 0.1 N/m to 50 N/m for the cantilevers typically used in ambient electromechanical measurements. In the second term of Eq. (1), we have substituted $A_{em} = d_{\text{eff}} V_{tip}$ into the expression, directly linking the electromechanical drive with the response amplitude. Cantilevers used in AFM applications are elastic beams that appear in a variety of shapes and sizes. The resonance frequency shift in response to tip-sample interactions is a strong function of the cantilever geometry.[41] Because they more closely obey analytic expressions, simple, diving-board geometries are best suited for quantitative measurements. However, there can be substantial variations in cantilever production process variations. It is possible for spring constants to vary by as much as an order of magnitude for cantilevers with the same nominal geometry. The cantilevers used in this work were silicon "ASYELEC-01" probes with nominal properties of $k_{cant} \approx 2$ N/m $and$ $\omega_0 \approx 2\pi \cdot 72$ kHz.[42]

The vast majority of current AFMs use a "split" photodetector to convert the motion of the cantilever into a measured voltage $V_{det}$.[43] In this optical beam detector (OBD) scheme, when the signal is close to null, the measured signal is roughly proportional to the slope (or bending) of the cantilever.[44,]

[45] Although presented in various ways, here we will use denote this proportionality constant the inverse optical lever sensitivity, $InvOLS$, where the amplitude in meters is related to the measured voltage through

$$A = InvOLS \cdot V_{det}. \tag{2}$$

$InvOLS$ can be estimated in a number of different ways, typically involving pressing the cantilever against a stiff, non-compliant surface a known distance while measuring $V_{det}$. The resulting data is fit to a line, and the slope yields $InvOLS$. In PFM measurements, this force curve calibration procedure then allows the piezo sensitivity $d_{\text{eff}}$ to be estimated by combining Eqs. (1) and (2).

### III. Frequency Dependent Imaging

To investigate the effect of frequency on PFM response, we performed imaging experiments on a commonly available sample, periodically poled lithium niobate (PPLN).[46] Lithium niobate (LiNbO$_3$) belongs to point group 3$m$, and therefore the piezoelectric matrix has four independent components: $d_{15}$, $d_{22}$, $d_{31}$, and $d_{33}$.[47] The vertical PFM signal far away from domain walls contains contributions from all four components, as described by Lei et al.,[48] but is dominated by the $d_{33}$ and $d_{15}$ contributions.

An example of frequency-dependent PFM on a PPLN sample is shown in Figure 1. Figures 1(a)-(f) show the PFM phase with the tip-sample potential oscillated at 40, 150, 190, 220, 310 and 330 kHz respectively, while Figures 1(a')-(f') show the corresponding PFM amplitude. Note that the last two frequencies were chosen to be just below and just above the contact resonance at ~320 kHz. The OBD spot was positioned at the tip, as is typical for maximizing the sensitivity of AFM measurements. The phase images in Figures 1(a), (e) and (f) show the expected 180° shifts over the two domains. However, in Figures 1(b), (c) and (d) the phase goes through a noisy reversal, inverting between 1(b) and (d). As seen in Figures 1(b') and (d'), the noisy phase images correspond to regions of small amplitude. The drive voltage amplitude was held constant at $V_{tip} = 3$ V during all of the measurements so the application of $d_{\text{eff}} = A_{em}/V_{tip}$ will lead to very different estimates of $d_{\text{eff}}$. Perhaps even more clearly, the amplitude in the central "down" stripe domain is different from that in the "up" domain and varies, appearing larger at low frequencies (see Figures 1(a') and (b')) and smaller at high frequencies (see Figure 1(c')-(f')).

The data of Figure 1 violates all three of the "ideal" PFM expectations listed above. In one previous study, asymmetries in the amplitude of the sort shown in Figure 1 were attributed to non-stoichiometric $LiNbO_3$[49] although this does not address either the lack of 180° phase shifts in Figure 1(b)-(d) or the frequency dependence of the amplitude measurements in Figure 1(a')-(f') . As we will discuss below, long-range electrostatic interactions, acting between the extended body of the cantilever and the surface can have significant and measurable effects on the cantilever mode shape and may explain asymmetries in the amplitude response.[50]

Figure 2(a) shows the phase spectra over the two domains at the red and black locations indicated in Figure 1(a), while Figure 2(b) shows the associated amplitude spectra. The amplitude axis is logarithmic, making the amplitude clearly visible at a broad range of frequencies despite the high quality factor ($Q \approx 130$) of the contact resonance. As with the data of Figure 1, these spectra were acquired with the OBD spot positioned at the tip of the cantilever. The frequencies used in the six images of Figure 1 are labeled as dashed vertical blue lines. One feature of note in the spectra for both the "up" and "down" domains is the existence of a minimum in the amplitude curve, at ~155 kHz for the "up" domain and ~225 kHz for the "down" domain. In both cases, there is a phase shift going through the frequency of minimum amplitude, despite being well below the contact resonance. (The 180° phase shift at the resonance frequency is a consequence of phase wrapping in the AFM lock-in amplifier.)

The red and black dashed lines in Figure 2(b) show SHO fits using the following parameters: $Q = 130, \omega_0 = 2\pi \cdot 310.49$ kHz, $V_{tip} = 3$ V, and $k_{cant} = 1$ N/m. In addition, the sensitivity of the cantilever was calibrated with the force curve technique to be $InvOLS = 100.1$ nm/V. With this calibration factor, the measured amplitude voltages were converted to values for the effective sensitivity $d_\text{eff}$ as discussed above, yielding estimates of $d_\text{eff} \approx 43$ pm/V (red, down domain) and $d_\text{eff} \approx 38$ pm/V (black, up domain). The estimates are significantly larger than the expected value of ~0.6-26 pm/V.[51,52] In addition, the values for the opposite domains differ by more than 10%, despite the amplitude independence expected in the ideal case. Perhaps most importantly, while the fits agree with the data near DC and resonance, there is significant disagreement at intermediate frequencies, where the fits do not predict the experimentally observed amplitude minima. Clearly the SHO model does not completely describe the measurements, motivating the mode mapping and modeling described and discussed below.

**IV. Mode Mapping**

Ernst Chladni is credited with mapping the modal shapes of plates in response to different boundary conditions.[53] The equivalent change of cantilever modal shape in response to changes in the boundary conditions is the fundamental basis for all dynamic AFM measurements. The mode shape changes are small in many measurement modes, but in other techniques such as contact resonance AFM, discussed theoretically by Rabe et al.,[42] the mode shape can experience radical changes. As part of their work, Rabe et al. mapped the mode shapes of a variety of *freely* vibrating rectangular levers using an optical interferometer. They found that the rectangular levers exhibited overall good agreement with cantilever beam theory, though they did observe some deviations they attributed to coupling between transverse and torsional modes. This initial work did not explore mode shapes of the cantilever tip interacting with the surface. Since that pioneering work, there have been a number of other studies of the variation in the dynamic mode shape of vibrating cantilevers.[54, 55, 56] In addition to these imaging applications, mode shape measurements have been useful for cantilever-based sensing.[57, 58]

To study frequency-dependent effects during PFM measurements in more detail, cantilever spectrograms – the response of the cantilever as a function of both frequency and OBD spot position – have been measured in situ while the cantilever probes a functional material. Figure 3(a) shows a schematic diagram of the setup used to scan the spot location over the cantilever while the lever is in contact with the surface.[59] We used the built-in motorized stage in a Cypher AFM to measure the cantilever deflection, amplitude and phase as a function of OBD spot location. The lateral positioning precision of ±200 nm and the vertical precision of ±40 nm of this system allows us to measure the response with great repeatability. The basic algorithm for making the measurements is as follows:

1. Engage the cantilever on the surface as you would for normal imaging or other measurements. This could be in contact mode, tapping or as the case for this report, PFM mode.
2. At a single x-y point on the surface, allow the z-position to stabilize with the z-feedback loop operating. If the z-position is drifting in time, this drift can be measured and used to extrapolate the control voltage when the z-feedback is disabled in the next step.
3. After stabilization or enough measurements to insure good extrapolation of the z-drift, disable the z-feedback. This is most easily accomplished by ramping the feedback gains to zero.
4. Reposition the OBD spot at various points on the cantilever while recording the OBD signal (deflection, amplitude and phase).
5. Sweep the drive frequency to record the amplitude and phase as a function of both position and frequency.
6. Return the spot to the original position and ramp the feedback gains to the original values. While doing this, the z-position as it responds to the feedback loop can be recorded to evaluate the effects of drift during the measurement steps 4-5. If there was significant drift, it is advisable to discard the measurements and attempt the experiment again.

Examples of spectrograms obtained in this manner are shown in Figure 4. Figures 4(a) and (a') show phase spectrograms for the down and up domains respectively, while Figures 4(b) and (b') show the corresponding amplitude spectrograms. The amplitudes were plotted using a logarithmic grayscale. The most notable features of the spectrograms are the nodal lines, visible as a swath of low-amplitude points in the amplitude images and as a 180° phase shift in the phase images (highlighted with dotted blue lines). An important point is that the nodal lines differ in their dependence on frequency and spot position for the two domains below resonance.

It is also clear from the phase and amplitude spectrograms that the crossover between the DC response and first contact mode, as indicated by the nodal lines, is measurable at surprisingly low frequencies. For the "down" domain in Figures 4(a) and (b), the crossover is on the order of half the contact resonance frequency. For the "up" domain in Figures 4(a') and (b'), the crossover is even lower – nearly to DC. The overlap region, where there is very little phase contrast in either domain, is shaded in Figure 4(c). Remarkably, the region lies in the middle of what has been conventionally recommended as the best operating region for PFM, specifically, (i) at a frequency well below the contact resonance[34] and (ii) with the OBD spot positioned close to the end of the lever.

**V. Theory**

Single point-mass models of the cantilever, even if long-range electrostatic interactions are considered, do not explain the behavior shown in Figures 1 and 2. Figures 4(d) and 4(d') show the SHO response calculated using Eq. (1). Since the cantilever mode shape is changing with frequency, the measured amplitude and phase at a given spot position will also change, leading to the clear disagreement between experiment amplitude spectrograms shown in Figures 4(b) and (b') and theoretical SHO amplitude spectrograms shown in Figures 4(d) and (d'). To account for this, it is necessary to go beyond the point-mass SHO model discussed above and consider the extended cantilever beam shape.

To model the extended beam shape, we use the Euler-Bernoulli (EB) model depicted schematically in Figure 3(b). After separating out the time dependence, the EB cantilever beam equation is given by

$$\frac{\partial^4 w(x)}{\partial x^4} - \beta^4 w(x) = \frac{3 F_{cant} L}{k_{cant}}, \tag{3}$$

where $\beta = C_1 \sqrt{\frac{\omega}{\omega_0}}$ and $L$ is the cantilever length. The delocalized electrostatic force $F_{cant}$ acts along the entire length of the cantilever and has been extensively discussed as a complication in interpreting PFM response.[60, 61, 62, 63, 64] Note that this term is divided by the cantilever stiffness $k_{cant}$, suggesting that the effect of long-ranged electrostatics can be reduced by choosing as stiff a cantilever as possible for a given measurement. Of course, there are practical limitations on $k_{cant}$, since a large loading force can damage tip coatings, the tip structure itself and the sample. On the clamped end of the cantilever, the displacement and slope (deflection angle) are zero: $w(0) = 0$ and $w'(0) = 0$. On the tip end, since we are neglecting any lateral component in the tip-sample interactions, $w''(L) = 0$. Localized forces, including both localized electrostatic interactions between the tip and sample along with localized piezoelectric strain forces, lead to the shear force boundary condition $w'''(L) = \frac{k_{ts}}{EI}(w(L) - z_{em})$, where the localized electromechanical strain is given by $z_{em} = d_{eff} V_{tip}$. $k_{ts}$ is the tip-sample contact stiffness and is related to the normalized contact stiffness α in Figure 3(b) by $\alpha = k_{ts}/k_{cant}$, while $E$ and $I$ are the cantilever's Young's modulus and cross-sectional area moment of inertia, respectively. At this point we should note that the amplitude measured by the OBD sensor is approximately the slope of the cantilever displacement: $A_{OBD} \approx w'(x)$. In the case where the displacement of the cantilever is measured directly, such as an interferometer, the detection position sensitive amplitude is equivalent to the displacement $A_{int} \approx w(x)$.[65]

Eq. (3) was solved analytically using Mathematica and then used to compute the theoretical EB spectrograms shown in Figures 4(e) and (e') and 5(a')-(c'). As discussed above, the slope $w'(x)$ of the theoretical EB spectrograms was calculated to approximate the OBD amplitude. While this approximation breaks down for small cantilevers and/or large laser spots, the nominally 3 μm long spot size combined with the 240 μm nominal cantilever length implies shape errors well under 1%.[44] We expect errors associated with deviations from an ideal EB diving board of uniform thickness to be a much larger source of error. The excellent agreement between the experimental and theoretical EB spectrograms below resonance justified omitting additional terms in the model such as the electrostatic coupling between the *tip* and sample (remembering that we are still keeping the delocalized electrostatic coupling between the cantilever body and the sample), the sample work function, the electrical contact quality or position-sensitive contact stiffness across the two domains.

At first look, the sub-resonance asymmetric nodal lines experimentally observed in Figures 4(b) and (b') are easily duplicated with this model, as shown in Figures 4(e) and (e') respectively. This asymmetry is consistent with a background electrostatic force due to electrostatic coupling between the body of the cantilever and the charged sample surface.[43]

The EB and SHO model calculations can be used to compare spectrograms made at applied DC bias voltages. The electrostatic coupling between the body of the cantilever and the sample surface was directly measured by positioning the cantilever ≲ 100 nm above the surface (out of contact) and then measuring the DC bias-dependent electromechanical response, keeping the AC amplitude at a constant 3 V ($V_{tip} = V_{DC} + 3V \cdot cos\omega t$). When $\omega = \omega_0$, we observed that the response amplitude was well described by $A = \gamma(|V_{DC} - V_{offset}|)$, where $\gamma$ and $V_{offset}$ were estimated as $\gamma \approx 12.1$ nm/V and $V_{offset} \approx -23.3$ mV (data not shown).[66] Assuming that the driving force for this non-contact experiment and the contact PFM measurements were similar allowed us to estimate the non-localized force as $F_{cant} = k_{cant}A/Q \approx 12.5$ nN/V · $|V_{DC}|$, where we have neglected the small DC offset. Figures 5(a)-(c) show bias-dependent spectrograms measured over a down domain made with $V_{DC} = +5$ V, 0 V and $-5$ V. The faint white vertical bands at 125 kHz in the experimental amplitudes correspond to an instrumental artifact and can be neglected. The bias-dependent evolution of the nodal lines (dark, low-amplitude regions) in the experimental amplitude measurements in Figures 5(a-c) is strikingly reproduced by the theoretical EB calculations shown in Figures 5(a'-c'), again in contrast to the theoretical SHO model results shown in Figures 5(a"-c"). Although beyond the scope of this work, this suggests it may be possible to control evolution of the nodal regions by actively controlling $V_{DC}$, in a manner similar to Kelvin Probe Force Microscopy (see, for example, ref. 71).

**VI. DISCUSSION**

These experiments were initially motivated by trying to improve quantitative PFM measurements, and the results indeed suggest strategies for improving these measurements. Notably, of the two independent variables in the spectrograms, the PFM response was most nearly ideal, with 180° phase shifts and equal amplitudes over opposite domains when the frequency was very low (<10 kHz). Near resonance, the phase contrast was well behaved, but there were differences in the amplitudes of the responses. While these differences may be explained with a slightly more complicated model than that presented in Eq. (3), it also implies that care needs to be taken with interpreting the response near resonance.[67, 68] At the same time, the enhanced signal to noise provided by imaging at resonance provides a benefit that can be well worth this tradeoff.

In the absence of making the careful position- and frequency-dependent measurements shown here, quantitative measurements can be improved by minimizing the effects of long-range coupling between the cantilever and sample in the following ways:

1. Choosing a low drive frequency. While this was indeed confirmed, the definition of "low" depends very strongly on the electrostatic term and may in some cases be well below even a few kHz.
2. If operating on resonance, which is desirable for improved signal to noise, care must be taken in interpreting the response. Specifically, changes in dissipation will change the quality factor and therefore the gain of the resonance amplifier.
3. Use of smaller cantilevers to reduce the electrostatic coupling between the tip and sample.
4. Use of longer tips, thus increasing the distance between the cantilever body and sample, reducing the capacitance.
5. Shielded probes. These may reduce the capacitance but are also more expensive and not as well developed as conventional cantilevers at this point.
6. Stiffer cantilevers also will reduce the effect of long-range electrostatic forces but may be undesirable for thin films and softer materials since the high loading force may damage the sample.
7. Positioning the OBD spot closer to the base of the cantilever can reduce the effect of nodal lines on phase and amplitude (at the cost of a reduction in sensitivity).
8. As pointed out by others,[69] scanning along the edge of a sample may help minimize these long-range electrical effects.
9. Measurements at different sample rotation angles could provide insight by varying the body-charge coupling.

**VII. CONCLUSION**

We have developed a method for measuring the dynamic response of a cantilever to tip-sample interactions while the cantilever is interacting with a sample. This technique can be applied to almost any AFM measurement mode, subject to the experimental constraints discussed in section IV above. These constraints still allowed systematic exploration of the PFM response of a cantilever scanning a PPLN sample. Recent developments in a combined OBD beam and interferometer AFM[70] allows modal mapping with the conventional feedback loop operational, bypassing some of the requirements for low drift and a time-independent tip-sample interaction that we had in this work. In the case of PFM measurements, we presented maps of the cantilever modal response as a function of OBD spot position and frequency through the first contact resonance. In particular, these measurements demonstrate that the conventional SHO model is insufficient to understand PFM response, at least for samples with a delocalized interaction between the sample and cantilever body such as the PPLN measured here.

By using an EB model of a PFM cantilever that includes a piezoelectric drive located at the tip and an extended electrostatic drive along the body of the cantilever, we have obtained excellent agreement between experimental and theoretical spectrograms. In the case where there are long-range electrical interactions between the cantilever and the sample, such as shown here, proper measurement of $d_{\text{eff}}$ depends on a choice of drive frequency and spot location. In particular, there are some locations where small changes in the boundary conditions (such as tip-sample stiffness) or the balance of electrostatic versus piezoelectric forces can cause contrast inversions and amplitude changes that may be mistaken for polarization changes in the sample. Note that these effects can occur at zero DC applied bias in the presence of even small surface charges.

Inspection of the spectrograms suggests that some of the worst effects associated with the frequency- and OBD-spot-dependent dynamics, including the movement of the nodal lines, can be minimized by positioning the OBD spot closer to the base rather than the tip of the cantilever. Although this results in a less sensitive OBD response, the benefit of a more stable response may be worth that sacrifice. Moving the spot closer to the base also widens the frequency range over which the response is well behaved. Finally, in the absence of making careful position- and frequency-dependent measurements of the type shown here, the effects of long-range coupling between the cantilever should be minimized with one or more of the following: smaller cantilevers, longer tips, shielded probes and scanning along the edge of a sample.

We anticipate that this approach to studying cantilever dynamics can be improved upon. One concern we have had is the potential presence of nonlinearities in the drive or detection method. An improvement we have started to explore is to operate in a constant amplitude mode, where the drive amplitude is modulated as a function of frequency to maintain the resulting cantilever motion at the same amplitude. In contrast to earlier mode measurement approaches, these measurements are fully integrated into the AFM. In addition, since the sensor is the same, the OBD sensitivity of the mode shape measurement exactly matches that of the AFM or PFM measurement. This can be run in any imaging mode including contact, contact resonance, PFM, ESM, tapping, dual AC and force-distance measurements, both at high and low speed.

**VIII. FIGURES**

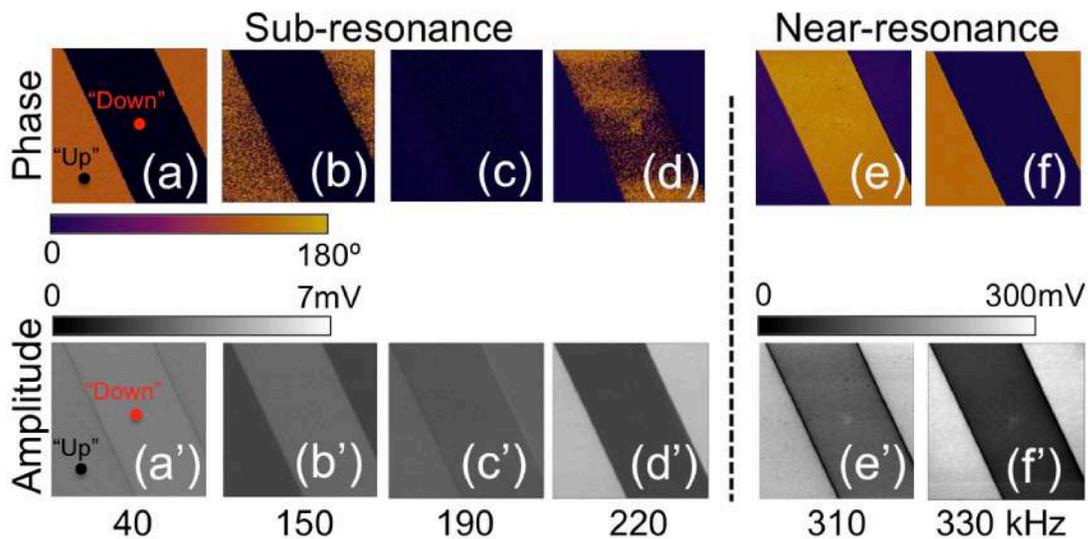

Figure 1. Frequency-dependent PFM response on a PPLN sample. The phase is shown in images (a)-(f) and the amplitude in images (a')-(f') for f = 40 kHz, 150 kHz, 190 kHz, 220 kHz and 330

kHz respectively. The contact resonance frequency was ~320 kHz. The 7 mV amplitude scale bar is for images (a')-(d'), and the 300 mV scale bar is for images (e') and (f'). The black and red dots indicate positions of up and down polarization respectively and were used for the measurements in Figure 2.

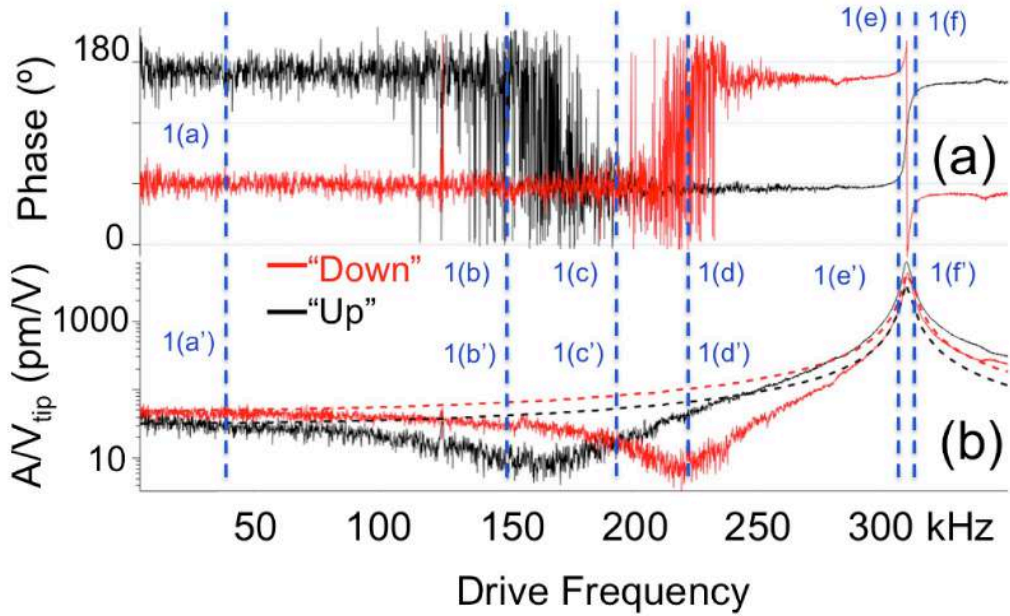

Figure 2. (a) Phase spectra over the two domains at the red and black locations indicated in Figures 1(a) and 1(a'). (b) Associated spectra of amplitude divided by drive voltage plotted on a log scale. The labels (a)-(d) and (a')-(d') indicate the drive frequencies used to acquire the images shown in Figure 1. The red and black dashed lines correspond to a SHO fit as described in the text. Note that while the fit agrees well near DC and resonance, there is significant disagreement at intermediate frequencies.

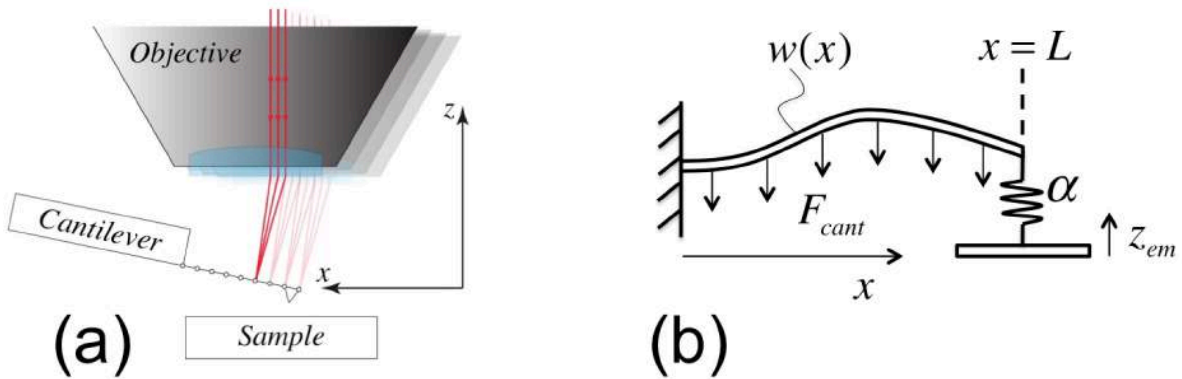

Figure 3. (a) Schematic of the motorized spot positioning system used to scan the OBD spot along the length of the cantilever. The spot can be scanned in all three dimensions, insuring that it remains in focus as it is scanned along the length of the tilted cantilever. (b) A schematic of the analytically modeled cantilever of length $L$ used to compute the theoretical spectrograms. $w(x)$ is the cantilever displacement along the length $x$. Long-range forces between the sample and the body of the cantilever are denoted by $F_{cant}$, while the localized electromechanical strain $z_{em}$ is coupled to the end of the cantilever ($x = L$) through a normalized tip-sample stiffness α.

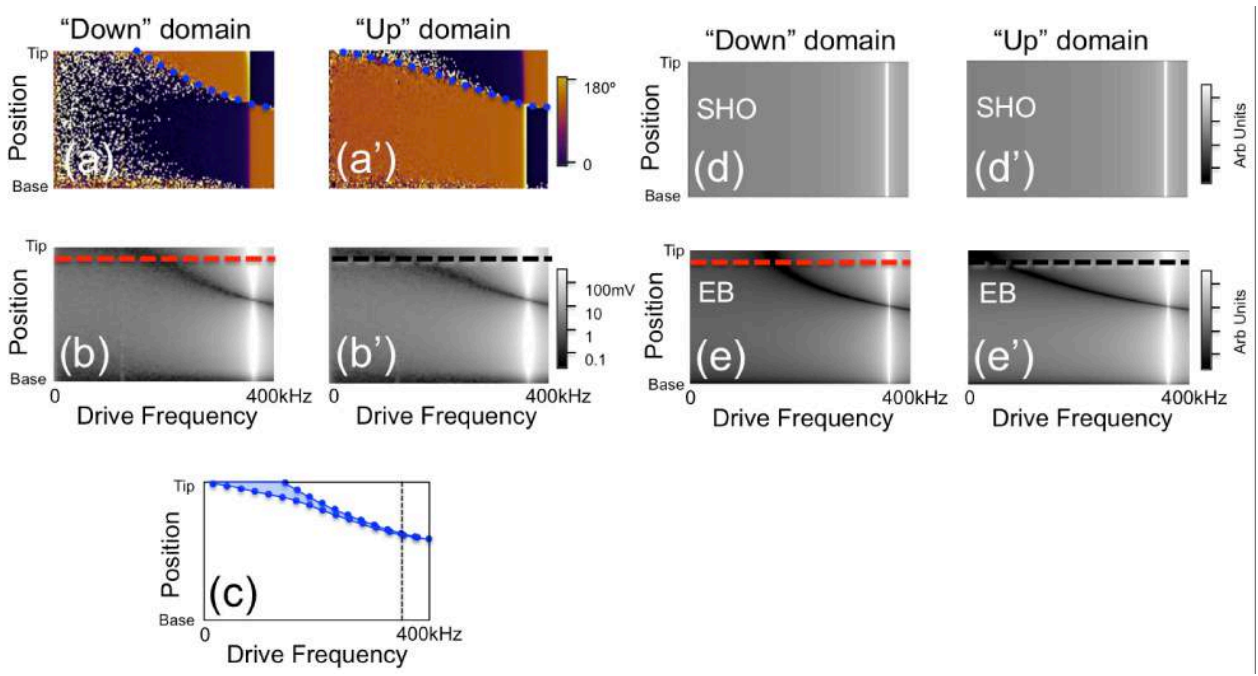

Figure 4. PFM spectrograms on a PPLN sample. (a) and (b) show the "down" domain phase and amplitude respectively, while the "up" domain phase and amplitude are shown in (a') and (b') respectively. The spectrograms show the experimental phase (a) and (a') and the logarithm of the experimental amplitude (b) and (b') as a function of frequency (0-400 kHz) on the horizontal axis and OBD spot position on the vertical axis, with the cantilever base on the bottom and the tip on the top. The dotted blue lines in (a) and (a') are guides to indicate the locations of the amplitude nodes, where the amplitude goes through a minima and the phase reverses. (c) shows the "trouble zone" region with roughly zero phase contrast shaded in blue. This region happens to be in the range where many

conventional sub-resonance PFM measurements take place, with the OBD spot positioned near the tip of the cantilever and the drive frequency at a value substantially lower than resonance. (d) and (d') show the OBD spot position-independent SHO amplitudes. Since the SHO model does not include any spot-position dependence, the nodes visible in the EB model and the experimental data do not appear. Finally, (e) and (e') show the theoretical EB model amplitude spectrogram over the "down" and "up" domains respectively. Specific parameters to obtain these theoretical results are discussed in the text. The dashed lines in (b), (b'), (e) and (e') indicate the OBD spot position near the end of the cantilever used to acquire the spectra in Figure 2.

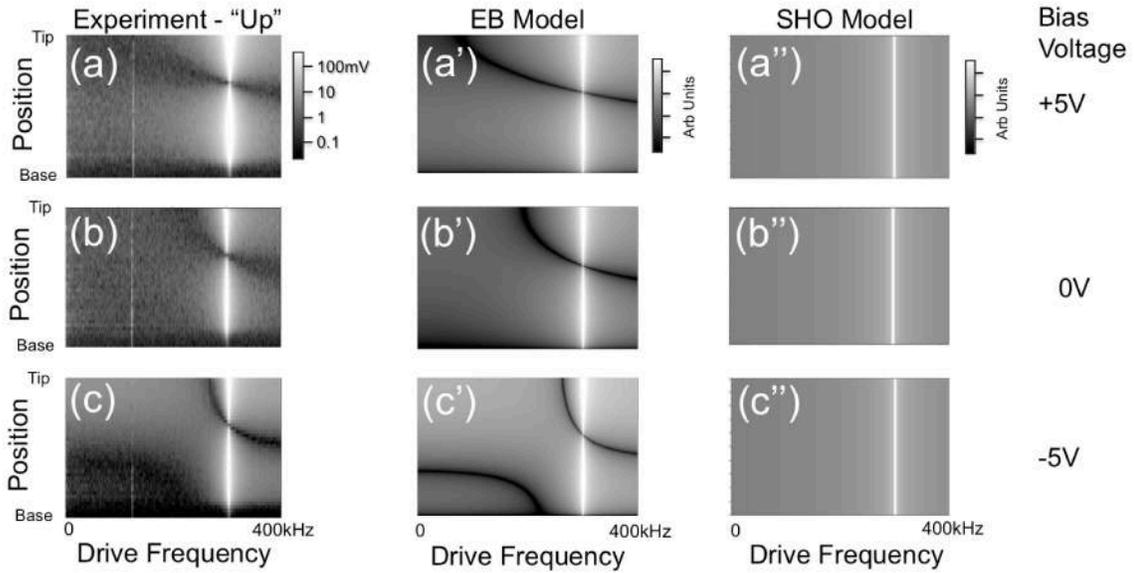

Figure 5. Experimental amplitude spectrograms using an ASYELEC-01 cantilever (see text) over a PPLN down domain for (a) +5V bias, (b) 0 V bias and (c) -5 V bias. (a'), (b') and (c') show the corresponding theoretical amplitude spectrograms for the respective bias voltages simulated with an EB cantilever model. (a''), (b'') and (c'') show the response predicted by the SHO model using the same parameters. Since the SHO model does not include spot-position changes, the nodes visible in the EB model and the experimental data do not appear.

IX. ACKNOWLEDGEMENTS

<>
The author thanks J. E. Sader for discussions and an introduction to beam theory within Mathematica™ during the early parts of this work. Ryan Wagner provided mathematical error checking and along with Arvind Raman and Donna Hurley and Ben Ohler, ongoing discussions, suggestions and edits. Elisabeth Soergel provided invaluable suggestions and lively and ongoing discussions.

## X. REFERENCES

[1] M. E. Lines and A. M. Glass, *Principles and Applications of Ferroelectrics and Related Materials*. Clarendon Press, Oxford, 1979.

[2] S. V. Kalinin, B. J. Rodriguez, J. D. Budai, S. Jesse, A. N. Morozovska, A. A. Bokov, and Z. G. Ye, Phys. Rev. B **81**, 064107 (2010).

[3] V. Shvartsman and A. Kholkin, J. Appl. Phys. **101**, 064108 (2007).

[4] V. Shvartsman, A. Kholkin, M. Tyunina and J. Levoska, Appl. Phys. Lett. **86**, 222907 (2005).

[5] A. Kholkin, A. Morozovska, D. Kiselev, I. Bdikin, B. Rodriguez, P. Wu, A. Bokov, Z. G. Ye, B. Dkhil and L. Q. Chen, Adv. Funct. Mater. **21**, 1977 (2011).

[6] G. A. Nazri and G. Pistoia, *Lithium Batteries: Science and Technology*. Springer-Verlag, New York, 2009.

[7] K. M. Abraham and Z. Jiang, J. Electrochem. Soc. **143**, 1 (1996).

[8] R. O'Hayre, S. W. Cha, W. Colella and F. B. Prinz, *Fuel Cell Fundamentals*. John Wiley & Sons, New York, 2009.

[9] S. V. Kalinin, B. J. Rodriguez, S. Jesse, T. Thundat and A. Gruverman, Appl. Phys. Lett. **87**, 053901 (2005).

[10] S. V. Kalinin, E. Eliseev and A. Morozovska, Appl. Phys. Lett. **88**, 232904 (2006).

[11] Y. M. Liu, Y.J. Wang, M. J. Chow, N. Q. Chen, F. Ma, Y Zhang and J. Li, Phys. Rev. Lett. **16**, 168101 (2013).

[12] G. Binnig, C. F. Quate and C. Gerber, Phys. Rev. Lett. **56**, 930 (1986).

[13] A. Gruverman and A. Kholkin, Rep. Prog. Phys. **69**, 2443 (2006).

[14] S. V. Kalinin, A.N. Morozovska, L. Q. Chen and B. J. Rodriguez, Rep. Prog. Phys. **73**, 056502 (2010).

[15] N. Balke, S. Jesse, A. N. Morozovska, E. Eliseev, D. W. Chung, Y. Kim, L. Adamczyk, R. E. Garcıa, N. Dudney and S.V. Kalinin, Nature Nanotech. **5**, 749 (2010).

[16] N. Balke, S. Jesse, Y. Kim, L. Adamczyk, A. Tselev, I. N. Ivanov, N. J. Dudney, and S. V. Kalinin, Nano Lett. **10**, 3420 (2010).

[17] K. Franke, H. Huelz and M. Weihnacht, Surf. Sci. **415**, 178 (1998).

[18] S. Hong, J. Woo, H. Shin, J. U. Jeon, Y. E. Park, E. Colla, N. Setter, E. Kim and K. No. J. Appl. Phys. **89** 1377 (2001).

[19] C. Harnagea, M. Alexe, D. Hesse and A. Pignolet, Appl.Phys. Lett. **83**, 338 (2003).

[20] B. D. Huey, C. Ramanujan, M. Bobji, J. Blendell, G. White, R. Szoszkiewicz and A. Kulik, J. Electroceram. **13**, 287 (2004).

[21] B. J. Rodriguez, C. Callahan, S. V. Kalinin and R. Proksch, Nanotechnology **18**, 475504 (2007).

[22] S. Jesse, S. Kalinin, R. Proksch, A. P. Baddorf and B. Rodriguez, Nanotechnology **18**, 435503 (2007).

[23] S. V. Kalinin, R. Shao, and D. A. Bonnell, J. Am. Ceram. Soc. **88**, 1077 (2005); S. Jesse, A. P. Baddorf and S. V. Kalinin, Nanotechnology **17**, 1615 (2006).

[24] S. Jesse, S. Guo, A. Kumar, B. J. Rodriguez, R. Proksch and S. V. Kalinin, Nanotechnology **21**, 405703 (2010).

[25] S. V. Kalinin, A. N. Morozovska, L. Q. Chen and B. J. Rodriguez, Rep. Prog. Phys. **73**, 056502 (2010).

[26] N. Balke, I. Bdikin, S. V. Kalinin, and A. L. Kholkin, J. Am. Ceram.Soc. **92**, 1629 (2009).

[27] S. V. Kalinin, A. Rar and S. Jesse, IEEE Trans. Ultrason. Ferroelectr. Freq. Control **53**, 2226 (2006).

[28] J. A. Christman, J. R. R. Woolcott, A. I. Kingon and R. J. Nemanich, Appl. Phys. Lett. **73**, 3851 (1998).

[29] B. J. Rodriguez, S. Jesse. Seal, A. P. Baddorf, S. Habelitz and S. V. Kalinin, Nanotechnology **20** 195701 (2009).

[30] H.-N. Lin, S.-H. Chen, S.-T. Ho, P.-R. Chen and I.-N. Lin, J. Vac. Sci. Technol. B **21**, 916 (2003).

[31] J. W. Burgess, J. Phys. D **8**, 283 (1975).

[32] C. Harnagea, A. Pignolet, M. Alexe, D. Hesse and U. Gösele, Appl. Phys. A: Mater. Sci. Process. **70**, 1 (2000).

[33] H. Bo, Y. Kan, X. Lu, Y. Liu, S. Peng, X. Wang, W. Cai, R. Xue and J. Zhu, J. Appl. Phys. **108**, 042003 (2010).

[34] S. Jesse, B. Mirman and S. V. Kalinin, Appl. Phys. Lett. **89**, 022906 (2006).

[35] C. Harnagea, M. Alexe, D. Hesse and A. Pignolet, Appl. Phys. Lett. **83**, 338 (2003).

[36] T. Jungk, A. Hoffmann and E. Soergel, Appl. Phys. Lett. **89**, 163507 (2006).

[37] T. Jungk, A. Hoffmann and E. Soergel, J. Microsc. **227**, 72 (2007).

[38] I. K. Bdikin, V. V. Shvartsman, S. H. Kim, J. M. Herrero and A. L. Kholkin, Mater. Res. Soc. Symp. Proc. **784**, 83 (2004).

[39] B. D. Huey, in *Nanoscale Phenomena in Ferroelectric Thin Films*, edited by S. Hong. Kluwer, New York, (2004).

[40] K. Seal, S. Jesse, B. J. Rodriguez, A. P. Baddorf and S. V. Kalinin, Appl. Phys. Lett. **91**, 232904 (2007).

[41] U. Rabe, K. Janser, and W. Arnold, Rev. Sci. Instrum. **67**, 3281 (1996).

[42] http://www.asylumresearch.com/Probe/ASYELEC-01,Asylum

[43] S. Alexander, L. Hellemans, O. Marti, J. Schneir, V. Elings, P. K. Hansma, M. Longmire and J. Gurley, J. Appl. Phys. **65**, 164 (1988).

[44] T. E. Schäffer and P. K. Hansma, J. Appl. Phys. **84**, 4661 (1998).

[45] T. E. Schäffer and H. Fuchs, J. Appl. Phys. **97**, 083524 (2005).

[46] AR-PPLN $LiNbO_3$ test sample, Asylum Research, Santa Barbara, CA: http://www.asylumresearch.com/Products/AR-PPLN/AR-PPLN.shtml

[47] M. Jazbinsek and M. Zgonik, Appl. Phys. B **74**, 407 (2002).

[48] S. Lei, E. A. Eliseev, A. N. Morozovska, R. C. Haislmaier, T. T. A. Lummen, W. Cao, S. V. Kalinin and V. Gopalan, Phys. Rev. B **86**, 134115 (2012).

[49] D. A. Scrymgeour and V. Gopalan, Phys. Rev. B **72**, 024103 (2005).

[50] S.V. Kalinin, D.A. Bonnell, Nano Lett. **4**, 555 (2004).

[51] A. W. Warner, M. Onoe and G. A. Coquin, J. Acoust. Soc. Am. **42**, 1223 (1967).

[52] R. T. Smith and F. S. Welsh, J. Appl. Phys. 42 2219 (1971).

[53] E. Chladni, *Entdeckungen über die Theorie des Klanges.* (1787).

[54] P. Vairac and B. Cretin, Surf. Interface Anal. **27**, 588 (1999).

[55] M. S. Allen, H. Sumali and P. C. Penegor, J. Dyn. Sys., Meas., Control **131**, 064501 (2009).

[56] M. Spletzer, A. Raman and R. Reifenberger, J. Micromech. Microeng. **20**, 085024 (2010).

[57] J. Mertens, M. Álvarez, and J. Tamayo, Appl. Phys. Lett. **87**, 234102 (2005).

[58] R. Wagner, J. P. Killgore, R. C. Tung, A. Raman and D. C. Hurley, Nanotechnology **26**, 45701 (2015).

[59] R. Wagner, A. Raman and R. Proksch, Appl. Phys. Lett. **103**, 263102 (2013).

[60] S. V. Kalinin and D. A. Bonnell, Phys. Rev. B **63**, 125411 (2001).

[61] T. Hochwitz, A. K. Henning, C. Levey, C. Daghlian, and J. Slinkman, J. Vac. Sci. Technol. B **14**, 457 (1996).

[62] S. Belaidi, P. Girard, and G. Leveque, J. Appl. Phys. **81**, 1023 (1997).

[63] S. Hong, J. Woo, H. Shin, J. U. Jeon, Y. E. Pak, E. L. Colla, N. Setter, E. Kim and K. No, J. Appl. Phys. **89**, 1377 (2001).

[64] S. Hong, H. Shin, J. Woo and K. No, Appl. Phys. Lett. **80**, 1453 (2002).

[65] A. Labuda and R. Proksch, Appl. Phys. Lett. **106**, 253103 (2015).

[66] See Fig. 2 in D. Ziegler, J. Rychen, N. Naujoks and A. Stemmer, Nanotechnology **18**, 225505 (2007).

[67] A. Gannepalli, D. G. Yablon, A. H. Tsou and R. Proksch, Nanotechnology **22**, 355705 (2011).

[68] S. Xie, A. Gannepalli, Q. N. Chen, Y. Liu, Y. Zhou, R. Proksch and J. Li, Nanoscale **4**, 408 (2012).

[69] S. Hong, S. Hyunjung Shin J. Woo and K. No, Appl. Phys. Lett. **80(8)** 1453 (2003).

24